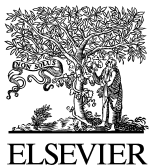

ELSEVIER

# High pressure phase transitions in BaWO$_4$

Vinod Panchal[a], Nandini Garg[a], A.K. Chauhan[b], Sangeeta[b], Surinder M. Sharma[a],*

[a]*Synchrotron Radiation Section, Bhabha Atomic Research Centre, Trombay, Mumbai 400085, India*
[b]*Technical Physics and Prototype Engineering Division, Bhabha Atomic Research Centre, Trombay, Mumbai 400085, India*

## Abstract

Using in-situ angle dispersive X-ray diffraction, we have shown that barium tungstate, which exists in scheelite phase at ambient conditions, transforms to a new phase at $\sim 7 \pm 0.3$ GPa. Analysis of our data based on Le Bail refinement suggests that this phase could be fergusonite and not HgMoO$_4$ type, which was proposed earlier from the Raman investigations. Beyond $\sim 14$ GPa, this compound undergoes another phase transformation to a significantly disordered structure. Both the phase transitions are found to be reversible.
© 2004 Elsevier Ltd. All rights reserved.

## 1. Introduction

The alkaline earth tungstates, molybdates and vanadates, aside from their geological relevance, are important laser host materials. Therefore, many of these ABO$_4$ type compounds (A = Ca, Ba, Bi, Sr…) (B = V, Mo, W) have been widely investigated [1–7]. Structurally these compounds exist in several forms, such as scheelite, wolframite, fergusonite, HgMoO$_4$, etc. [2,8–11]. Though earlier arguments, based either on packing efficiency or higher coordination, etc. [12], suggested wolframite to be a preferred high pressure structure, subsequent experiments have shown a wide variety of structures in the phase diagrams of these compounds. This necessitates a fresh look into some of the earlier studies. BaWO$_4$ is one such compound. At ambient conditions it is known to exist in the scheelite structure, shown in Fig. 1 (tetragonal, $I4_1/a$, $Z = 4$). Several recent high pressure studies have shown that the compounds having initial scheelite structure may transform to wolframite [3,13,14], fergusonite [10], HgMoO$_4$ [5], or other closely related structures. However, previous X-ray diffraction studies on barium tungstate, quenched from 4 GPa and 600 °C, indicated that it transforms irreversibly to a monoclinic phase ($P2_1/n$), which is distinct from either wolframite or fergusonite structures [15, 16]. Subsequent high pressure Raman scattering by Jayaraman et al. [5], showed that barium tungstate undergoes a reversible first order phase transition around $6.5 \pm 0.3$ GPa. From the observed abrupt decrease in the frequency of the internal W–O modes, these authors suggested the high pressure structure to be HgMoO$_4$ type [11]. However, no in-situ X-ray diffraction study has been carried out on this compound. Hence, we have carried out angle-dispersive X-ray diffraction studies on barium tungstate under high pressures, upto 20 GPa.

## 2. Experimental

High pressure angle dispersive powder X-ray diffraction experiments were carried out employing monochromatized (with graphite (002) monochromator) molybdenum K$_\alpha$ radiation from a rotating anode X-ray generator. Finely powdered sample was loaded in a pre-indented steel gasket hole of $\sim 150$ μm diameter in a Mao-Bell kind of diamond anvil cell. For these experiments either 4:1 methanol–ethanol or 16:3:1 methanol–ethanol–water mixtures were used as pressure transmitters, which are known to freeze into soft solids at ∼10.4 and ∼14.5 GPa, respectively [17,18]. The pressure in the gasket hole was determined by monitoring the shift in ruby R-lines [19]. At each pressure the diffraction data was collected for more than 15 h on an imaging plate (IP). One-dimensional diffraction profiles were generated from IP records using FIT2D software [20].

* Corresponding author. Fax: +91-22-5505341.
*E-mail address:* smsharma@magnum.barc.ernet.in (S.M. Sharma).

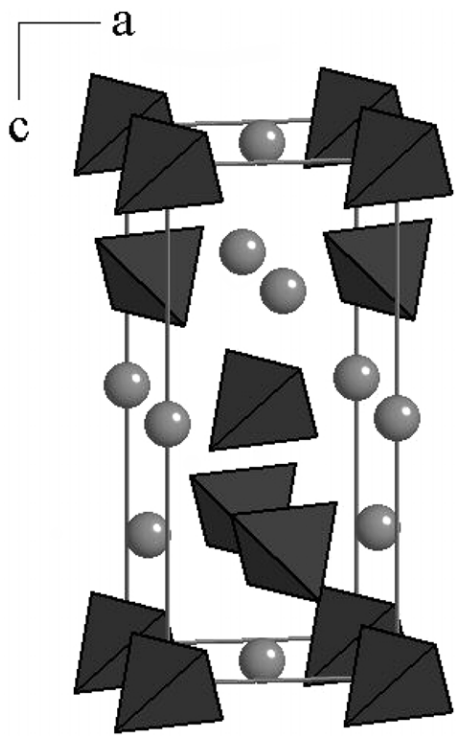

Fig. 1. Crystal structure of BaWO₄ at ambient conditions (scheelite). The W atoms are tetrahedrally coordinated to oxygen and ⬤ represent Ba.

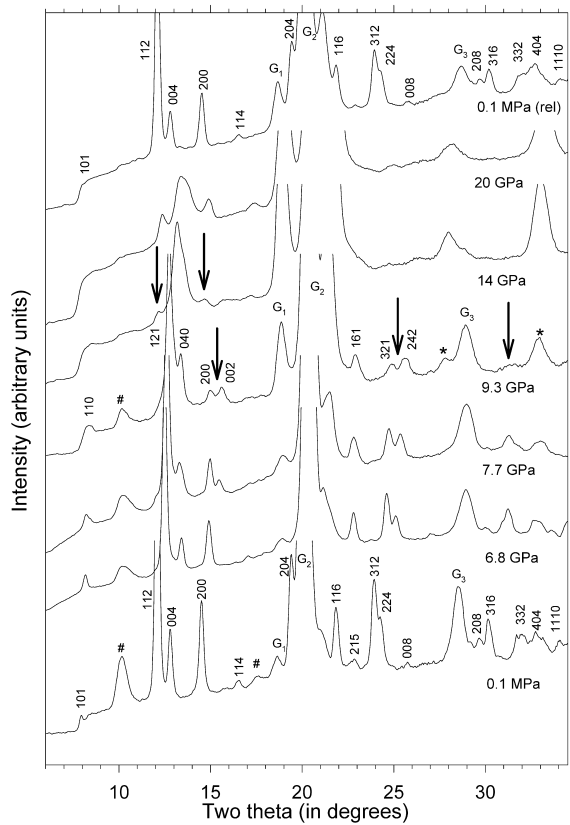

Fig. 2. The diffraction patterns of BaWO₄ at various pressures. The diffraction pattern at 0.1 MPa has been indexed to the scheelite phase and the pattern at 9.3 GPa to the fergusonite phase. The arrows mark the observed changes in the diffraction pattern at the completion of the first phase transformation (9.3 GPa), and after the second phase transformation (14 GPa). $G_i$ ($i = 1$–3) indicate the gasket peaks. The diffraction peak at 10° (#) is the second order (200) diffraction peak of the steel gasket. The gasket steel is primarily the fcc phase. However, it has a small amount of some other phase, which shows up on either side of the $G_3$ (200) gasket diffraction peak marked as (*). Beyond 10 GPa the gasket undergoes a phase transformation and the diffraction peaks of the fcc phase of the gasket weakens and the intensities of the diffraction peaks marked with the * increase. The diffraction from the gasket is quite strong as the X-ray beam from the rotating anode generator is a diverging beam.

## 3. Results and discussions

Fig. 2 shows our recorded X-ray diffraction profiles at a few representative pressures. Due to the divergence of the X-ray beam, the recorded diffraction patterns are contaminated by several diffraction peaks from the gasket (marked as $G_i$ in Fig. 2). However, at ambient conditions, 14 diffraction peaks of scheelite–BaWO₄ are clearly discernible and the cell constants for the initial phase are found to be, $a = 5.63 \pm 0.01$ Å and $c = 12.75 \pm 0.01$ Å; which are in fairly good agreement with the published results—$a = 5.61$ Å, $c = 12.71$ Å (ICDD card no 43-0646). Our results show that upto ∼7 GPa, the observed diffraction peaks shift smoothly with pressure. However, at ∼7 ± 0.3 GPa, an additional diffraction peak emerges close to the (200) diffraction peak (marked with an arrow in Fig. 2). Concomitantly, several diffraction peaks broaden and the intensities of the (312) and (224) peaks redistribute. These observed changes suggest a phase transformation to a lower symmetry structure—broadly supporting the earlier Raman studies of Jayaraman et al. [5]. In addition, the observed coexistence of the scheelite (phase I) and the new phase (phase II) upto 7.7 ± 0.3 GPa, indicate the first order nature of this transformation. On further increase of the pressure, signal to noise ratio of the diffraction peaks deteriorates significantly beyond 10 GPa. However, at ~ 14 GPa, two new diffraction peaks emerge, one on either side of the diffraction peak at $2\theta \sim 13°$. These new peaks, indicated by arrows in Fig. 2, persist upto ~ 20 GPa, the highest applied pressure in these experiments. From the limited diffraction data available beyond 14 GPa, it is difficult to identify the underlying structural modifications. However, it is also possible that BaWO$_4$ transforms to a disordered crystalline phase (in the sense that amorphous and a poorly crystallized phases co-exist, similar to what has been observed in $\alpha$-FePO$_4$ [21] and $\alpha$-AlPO$_4$ [22]). Since the diffraction peaks remain broad even on the release of pressure below 10 GPa, it is more likely that the broadening and degradation of diffraction pattern is due to a transformation to a disordered crystalline phase. It was observed that the experimental results did not vary with the usage of different pressure transmitters.

Earlier high pressure studies of the scheelite tungstates suggest that the probable space groups of the phase II may be—$I2/a$ (HgMoO$_4$ or fergusonite), $P2/c$ (wolframite) or $P2_1/n$. The computed diffraction pattern for the $P2_1/n$ space group differs considerably from the observed diffraction pattern of the phase II in terms of positions and intensities of the peaks, and is thus ruled out. Also, a strong diffraction peak of the wolframite structure, expected at $d \approx 4$ Å, is not observed in the phase II. In addition, the intensities of the observed diffraction peaks at $d = 2.67$ and $2.59$ Å differ considerably from the calculated intensities for the wolframite structure. Moreover, as given in Table 1, the observed diffraction peak at $d = 1.78$ Å and a doublet, observed between $1.59$ Å $\leq d \leq 1.64$ Å, are not ascribable to the wolframite structure (the peak close to ~ $d = 1.59$ Å would be very broad). These disagreements with the observed results do not support wolframite as a favorable structure of phase II. To evaluate HgMoO$_4$ and fergusonite structures (space group $I2/a$), we have analyzed our data with Le Bail fitting [23]. Background corrected diffraction pattern could be reasonably well fitted with the Le Bail method (with $R_{wp} = 0.08$), supporting this assignment of the space group. Further, the comparison of the observed diffraction pattern and the calculated diffraction patterns of the fergusonite phase and HgMoO$_4$ phase, as presented in Table 1, shows that the intensity as well as the $d$-value of the first diffraction peak of the HgMoO$_4$ structure differs substantially from the observed value. Also we do not observe a strong (200) peak of the HgMoO$_4$ phase. However, for the fergusonite structure, the intensities as well as the $d$-values match fairly well. Based on these observations, we conclude that the observed diffraction pattern of phase II favors fergusonite structure in comparison with the wolframite and HgMoO$_4$ structures. Fig. 3 shows the BaWO$_4$ in the fergusonite phase, generated by using the fractional atomic coordinates of Ref. [10]. In this phase though the W–O are still four coordinated, the WO$_4$ tetrahedra are significantly deformed. However, ideally the fractional coordinates should be determined by Rietveld analysis, which has not been carried out due to the contamination of diffraction pattern by the gasket peaks. Further, the comparison of Figs. 1 and 3 shows that the two structures are easily relatable through correlated-displacive atomic motion, i.e. the lattices of alternate layers (along $c$-axis) of the scheelite structure shift in opposite directions to form pairs with neighbor layers.

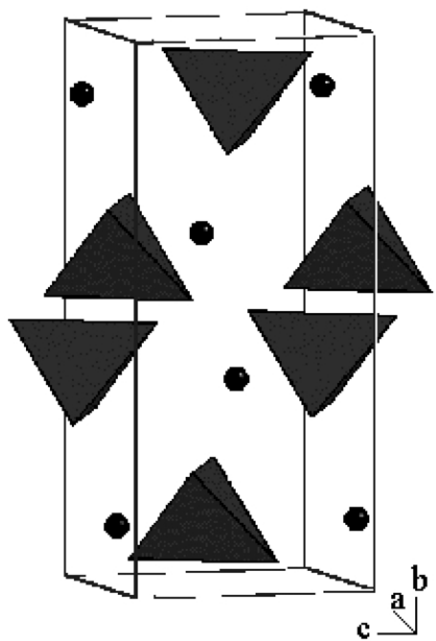

Fig. 3. Crystal structure of BaWO$_4$ in the fergusonite phase (with the fractional atomic coordinates of Ref. [10]). The W atoms are tetrahedrally coordinated to oxygen and ● represent Ba.

The observed variation of the lattice constants $a$ and $c$ with pressure is shown in Fig. 4(a)–(c). The solid line is a quadratic fit, which gives

$$a(p) = 5.6257 - 0.0276p + 6.929 \times 10^{-4}p^2 \quad (1)$$

$$c(p) = 12.749 - 0.0987p + 2.002 \times 10^{-3}p^2 \quad (2)$$

where $p$ is in GPa.

Experimentally determined equation of state, plotted in Fig. 4(d), shows a small volume discontinuity (1.7%) at ~7 ± 0.3 GPa. The $P$–$V$ data can be analyzed with the help of Birch–Murnaghan (BM) equation of state, viz.

$$P = (3/2)K[(V_0/V)^{7/3} - (V_0/V)^{5/3}][1 - (3/4)(4 - K')\{(V_0/V)^{2/3} - 1\}] \quad (3)$$

Where $K$ and $K'$ represent the bulk modulus and its derivative. A least square fit of our data with Eq. (3) gives $K = 57$ GPa and $K' = 3.5$ for the scheelite phase.

The anisotropy in compression, as discussed by Angel [24], is also analyzable with the help of the linear variant of BM equation of state. To do so, we can replace $V_0/V$ in Eq. (3) with $(a_0/a)^3$ or $(c_0/c)^3$, respectively, for $a$ and $c$ cell

Table 1

X-ray diffraction data for BaWO$_4$ at 8.3 GPa—the comparison of the observed d—spacings and intensities of phase II with the calculated ones from different competing structures

| Present study at 8.3 GPa | | Fergusonite [10][a] | | | Wolframite [2][b] | | | HgMoO$_4$ [11][c] | | |
|---|---|---|---|---|---|---|---|---|---|---|
| $d_{obs}$ | $I_{obs}$[d] | (hkl) | $d_{cal}$ | $I_{cal}$ | (hkl) | $d_{cal}$ | $I_{cal}$ | (hkl) | $d_{cal}$ | $I_{cal}$ |
| 4.92 | W | (110) | 4.98 | 4 | (100) | 5.16 | 6 | (011) | 5.36 | 21 |
|  |  |  |  |  |  |  |  | (002) | 5.36 | 9 |
| 4.76 | W | (011) | 4.8 | 2 | – | – | – | – | – | – |
| 4.1 | Second order gasket |  | – |  | (011) | 4.1 | 10 | (110) | 4 | 5 |
|  |  |  |  |  | (110) | 4 | 57 |  |  |  |
| 3.19 | VVS | (121) | 3.22 | 100 | (111) | 3.21 | 100 | (112) | 3.22 | 100 |
|  |  | (−121) | 3.21 | 98 | (−111) | 3.2 | 95 | (−112) | 3.21 | 99 |
| 3.03 | VS | (040) | 3.07 | 33 | (020) | 3.12 | 14.6 | (020) | 3.1 | 32 |
| 2.67 | M | (200) | 2.72 | 25 | (002) | 2.72 | 34 | (004) | 2.68 | 23 |
|  |  |  |  |  | (021) | 2.71 | 49 |  |  |  |
| – | – | – | – | – | – | – | – | (200) | 2.64 | 30 |
| 2.59 | M | (002) | 2.62 | 24 | (200) | 2.58 | 24 | (−121) | 2.59 | 11 |
| – | – | (150) | 2.24 | 3 | (112) | 2.25 | 14 | – | – | – |
|  |  |  |  |  | (−112) | 2.24 | 10 |  |  |  |
| – | Under gasket | – | – | – | – | – | – | (123) | 2.14 | 14 |
| – | Under gasket | (240) | 2.04 | 35 | (022) | 2.05 | 12 | (024) | 2.03 | 24 |
|  |  |  |  |  |  |  |  | (220) | 2.01 | 24 |
| – | Under gasket | (042) | 2 | 33 | (220) | 1.99 | 11 | – | – | – |
| – | Under gasket | – | – | – | (130) | 1.9 | 34 | – | – | – |
| – | Under gasket | (202) | 1.89 | 13 | (202) | 1.89 | 19 | (204) | 1.89 | 16 |
|  |  | (−202) | 1.88 | 16 | (221) | 1.87 | 29 | (−204) | 1.87 | 12 |
|  |  |  |  |  | (−221) | 1.86 | 26 |  |  |  |
|  |  |  |  |  | (−202) | 1.86 | 19 |  |  |  |
| 1.78 | S | (161) | 1.80 | 22 | (212) | 1.8 | 0.05 | (132) | 1.81 | 12 |
|  |  | (−161) | 1.799 | 20 |  |  |  | (−132) | 1.80 | 9 |
|  |  |  |  |  |  |  |  | (033) | 1.79 | 9 |
| 1.64 | M | (321) | 1.66 | 19 | (113) | 1.65 | 15 | (116) | 1.64 | 16 |
|  |  | (−321) | 1.65 | 19 | (−113) | 1.64 | 20 | (−116) | 1.63 | 17 |
| 1.59 | M | (242) | 1.61 | 15 | (311) | 1.59 | 15 | (224) | 1.61 | 11 |
|  |  | (123) | 1.61 | 19 | (−311) | 1.58 | 15 | (−312) | 1.61 | 14 |
|  |  | (−242) | 1.60 | 14 | (132) | 1.57 | 19 | (−224) | 1.61 | 12 |
|  |  | (−123) | 1.60 | 15 | (−132) | 1.566 | 17 |  |  |  |
|  |  |  |  |  | (023) | 1.56 | 18 |  |  |  |

The volume per formula unit of all the three structures was taken to be equivalent to the observed volume at 8.3 GPa. Since the beta angle is not unique for a particular structure (i.e. it varies from 89.6° [9] to 95.2° [10] for the fergusonite phase; from 89.4° [2] to 94.85° [14] for the wolframite phase etc) we have taken the beta angle equal to 89.6° (Le Bail fit of present study) for all the three structures. The fractional coordinates were taken from the references mentioned above in each column.

[a] Calculated values are from a (fergusonite) monoclinic unit cell with $a = 5.4444$ Å, $b = 12.2903$ Å, $c = 5.2358$ Å, $\beta = 89.56°$, $Z = 8$ and the fractional coordinates from Ref. [10].

[b] For wolframite, the calculations were carried with $a = 5.1643$ Å, $b = 6.2464$ Å, $c = 5.4303$ Å, $\beta = 89.56°$, $Z = 4$ and the fractional coordinates from Ref. [2].

[c] The calculations for HgMoO$_4$ structure were carried out with $a = 5.2749$ Å, $b = 6.194$ Å, $c = 10.7226$ Å, $\beta = 89.56°$, $Z = 8$ and the fractional coordinates from Ref. [11].

[d] VVS—very very strong, VS—very strong, S—strong, M—medium, W—weak.

lengths. K obtained by fitting this isothermal equation is then one third of the inverse of the linear zero pressure compressibility ($\beta_0$) along the chosen axis (where $\beta_0 = l_0^{-1}$ $(\partial l/\partial P)_{p=0}$, $l_0$ being the cell length at ambient conditions) [24]. For scheelite phase the observed variations in cell parameters (Figs. 4(a)–(c)) give the modulus $K_c$ (i.e., along the c-axis) to be 113 GPa, and $K_a$ (the modulus along the a-axis) to be a-axis) to be 197 GPa. This implies that in the scheelite phase c-axis is almost 1.7 times more compressible than the a-axis. The overall bulk modulus when calculated by the following equation

$$1/K = 2/K_a + 1/K_c$$

is 53 GPa which is slightly less than that calculated by fitting the $P-V$ data. The compressibility in this material is primarily due to the larger and softer Ba–O bonds (2.78 Å) rather than the shorter and stronger W–O tetrahedral bonds (1.87 Å). Though no direct information is available for BaWO₄, but for a closely related scheelite SrWO₄, the calculated bond stretching and bond bending force constants of W–O are about five times larger than that of Sr–O [6].

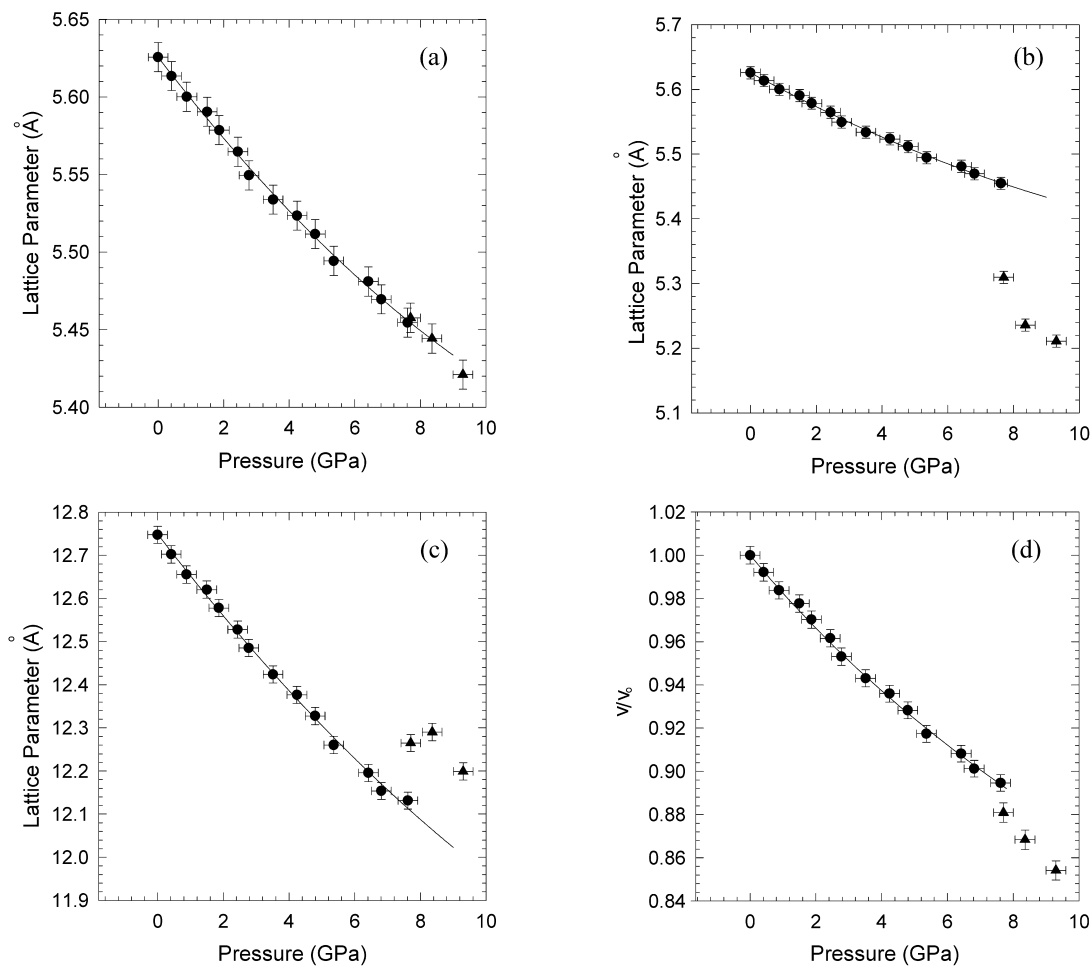

Fig. 4. Pressure dependence of lattice parameters and equation of state for the scheelite and the fergusonite phases of BaWO₄. The solid line in Fig. 4(a)–(c) shows the quadratic fit to the respective lattice parameters upto 7.6 GPa. (a) Lattice parameter $a$ for the scheelite phase (●) and for the fergusonite phase (▲). (b) Lattice parameter $b$ for the scheelite phase (●) and $c$ for the fergusonite phase (▲). (c) Lattice parameter $c$ for the scheelite phase (●) and $b$ for the fergusonite phase (▲). (d) Equation of state $\sim$10 GPa (●) for the scheelite phase and (▲) for the fergusonite phase. The solid line shows the fit to Birch–Murnaghan equation of state.

On release of pressure from 20 GPa, both the high pressure phases continue to exist upto $\sim$3 GPa, indicating a large hysteresis. The existence of hysteresis as well as reversibility between scheelite, fergusonite and disordered crystalline phase imply the first order as well as displacive nature of these phase transformations. It is interesting to note that similar, but temperature induced, reversibility between these two phases has also been observed in BiVO₄, rare earth orthoniobates and orthotantalates [25–28].

## 4. Conclusions

In-situ high pressure powder X-ray diffraction studies on BaWO₄ show that barium tungstate undergoes a first order phase transition to a new phase at $\sim$7 ± 0.3 GPa. It has been shown that this new phase could have a structure similar to the monoclinic fergusonite phase. At 14 GPa, it further transforms to a new phase which becomes significantly disordered by $\sim$20 GPa. On release of pressure from 20 GPa, the scheelite phase is recovered below 3 GPa, indicating reversible nature of both the transformations. As, in several ABO₄ compounds, similar scheelite-fergusonite phase transformation have been found to be ferro-elastic, it will be interesting to study whether this pressure induced transformation is also driven by a softening of some elastic constants.